# Distance, Borders, and Time: The Diffusion and Permeability of Political Violence in North and West Africa

D.B. Skillicorn[1], O. Walther[2], Q. Zheng[1], C. Leuprecht[3]

Abstract: This paper explores the spatial and temporal diffusion of political violence in North and West Africa. It does so by endeavoring to represent the mental landscape that lives in the back of a group leader's mind as he contemplates strategic targeting. We assume that this representation is a combination of the physical geography of the target environment, and the mental and physical cost of following a seemingly random pattern of attacks. Focusing on the distance and time between attacks and taking into consideration the transaction costs that state boundaries impose, we wish to understand what constrains a group leader to attack at a location other than the one that would seem to yield the greatest overt payoff. By its very nature, the research problem defies the collection of a full set of structural data. Instead, we leverage functional data from the Armed Conflict Location and Event Data project (ACLED) dataset that, inter alia, meticulously catalogues violent extremist incidents in North and West Africa since 1997, to generate a network whose nodes are administrative regions. These nodes are connected by edges of qualitatively different types: undirected edges representing geographic distance, undirected edges representing borders, and directed edges representing consecutive attacks by the same group at the two endpoints. We analyze the resulting network using novel spectral embedding techniques that are able to account fully for the different types of edges. The result is a "map" of North and West Africa that depicts the permeability to violence. A better understanding of how location, time, and borders condition attacks enables planning, prepositioning, and response.

[1] School of Computing, Queen's University, Kingston, Canada (skill@cs.queensu.ca);
[2] Department of Political Science, University of Southern Denmark, Sønderborg, Denmark and Division of Global Affairs, Rutgers – The State University of New Jersey, Newark, United States.
[3] Department of Political Science, Royal Military College of Canada, Kingston, Canada.

# 1. Introduction

The study of how crime and political violence diffuse across time and space has greatly benefited from the increasing availability of geo-referenced data and the use of spatial statistical analysis (O'Loughlin and Raleigh, 2008; Zammit-Mangion *et al.* 2013). In urban policing, for example, the design and use of hot-spot analysis based on historical data allows to predict when and where various kinds of crime are most likely to occur, and prepositioning policing assets accordingly (Braga, 2005). In this limited sense, predictive modelling of crimes has been remarkably effective. The urban environment lends itself to this kind of analysis: criminals are creatures of habit, they tend to travel limited distances, and some areas are naturally more target-rich than others.

If we try to adapt this approach to attacks by Violent Political Organizations (VPOs) in North and West Africa, there are some obvious difficulties. Just as in urban settings, some natural targets attract repeated attacks, for example, foreign workers in West African capitals or government forces stationed on military bases. Most victims of recent conflicts in the region are, however, civilians, killed in a rather unpredictable manner by armed groups whose main objective is ethnic or tribal homogeneity (Kaldor 2012). In such 'new wars', control over people, not territory, leads a multiplicity of state and non-state actors to build a complex ecosystem of affiliated and opposing groups that also constrain when and where an attack by a particular group might take place (Walther and Tisseron 2015; Zheng *et al.* 2015). Attacks also reflect competition between traffickers and violent extremist groups struggling to control Trans-Saharan criminal networks, and who often clash far from inhabited areas (Lacher 2012). Furthermore, many violent groups in the region do not limit their attacks to a particular 'turf' as urban gangs might; instead, they move relatively freely across the region, including across state boundaries.

The situation is far removed from a conventional Clausewitzian framework: attack the enemy with maximum force at its strongest point. Insurgent groups have fewer resources and compensate by striking at locations that maximize impact, even abstractly via publicity, while minimizing cost. Attackers often rely on guerilla warfare. They avoid head-on confrontation, thereby blurring the line between zones of war and zones of peace. Naval battles are a more apt analogy. Notwithstanding strategic constraints (such as the need to blockade enemy fleets at Trafalgar and Jutland), the precise locations at which battles occur are not contingent on terrain in the way in which many engagements on land are.

This paper explores the spatial and temporal diffusion of political violence in North and West Africa. To this end, it models the strategic landscape in a group commander's mind. The location of an attack requires a complex calculus that combines properties of the comparative attractiveness of targets, the physical geography of the terrain between the current location and potential targets, the obstacles and impediments to movement between the current location and targets, including borders that must be crossed, the difficulty of operating close to targets, and the need to maintain an element of surprise. We wish to understand what motivates or constrains a group leader to attack at a location other than the most obvious one (that is, the one that would seem to yield the greatest overt payoff).

The paper leverages the Armed Conflict Location and Event Data project (ACLED) dataset that catalogues violent extremist incidents in North and West Africa since 1997. We use this data to generate a novel form of "social network" whose nodes are administrative regions, and whose edges are of qualitatively different types": undirected edges representing geographic distance, undirected edges representing the impact of having to cross borders, and directed edges representing consecutive attacks by the same group at two locations. We analyze the resulting network using novel spectral embedding techniques that combines these different edge types into a "map" of North and West Africa that depicts the permeability to attacks from the perspective of a violent group. When distributions are highly skewed, statistical measures such as averages are useless for planning effective counterinsurgency deployment. This map of permeability reflects the impact of distance, borders, and time and violent group actions, and so provides a first step towards principled planning, prepositioning, and response.

The article proceeds as follows. The next section outlines existing literature on the two geographic features that are most likely to influence how attacks are conducted across space and time: the distance between places, and the impediment of state boundaries. Section 3 describes the geographical and temporal distribution of attacks in the region. Section 4 develops novel spectral techniques that model the effects of border costs and analyzes attack location over time. Section 5 discusses the main implications of our work before concluding.

## 2. Networks, space and borders

Over the last decade, geographers and network scientists have opened several avenues of research regarding the spatiality of social networks, i.e. how location, spatial practices or representations, and geographic arrangements of networks influence social ties (Carrasco *et al.* 2008; Adams *et al.* 2011; Schaefer 2011; Arentze *et al.* 2012). Space is now widely recognized as a fundamental dimension of politically violent organizations that often conduct operations from a territorial base, leverage geographic havens, compete with sovereign states, and fight for control over aspirational homelands (Medina and Hepner, 2013). As a result, an increasing number of scholars are working to integrate social network analysis and spatial analytical techniques (Carley and Pfeffer, 2012; Gelernter and Carley, 2015). As Carley (2006: 3) argues: 'If we look only at the social network then the focus of attention is on hierarchies, communication and other social relations. The addition of events and locations facilitates course of action analysis and enables linkage to various strategic planning tools'. Recent conceptual and technical developments related to the spatiality of social networks have primarily been applied to case studies located in the U.S., Middle East, Afghanistan and Pakistan, and Southeast Asia (Berrebi and Lakdawalla, 2007; Hannigan et al. 2013; Gao *et al.* 2013) or at the global level (Townsley et al. 2008, Medina and Hepner, 2011). By contrast, North and West Africa have received little attention from network science.

While existing research has made a major contribution to understanding the way in which violence emerges and diffuses over time and space, most of the studies conducted so far have apprehended the spatiality of social networks based on actors for whom the location or the territory was well known. In Radil *et al.*'s (2010), for example, urban gangs produces distinct spatial patterns depending on whether the groups are close geographically or topologically. In their study of the geography of the First World War, Flint *et al.* (2009) adopt a similar perspective by showing how the geographical and topological position of nation-states helps explain their power relations as a function of alliances and rivalries. In both cases, social networks that bound the attackers and their victims are assumed to be linked to a specific location or territory: gangs in Los Angeles delineate their turf through violence, while aggressions during the First World War were related to state territories. Building on very mobile violent organizations, this article seeks to address this gap by considering the effect of two fundamental geographic features on their movements across time and space: distance and borders.

The common assumption that space shapes social relations is often based on physical distance. Proximity increases network density by increasing the propensity to have contact with other social actors, which enhances individual integration, cohesion and shared values (Wellman, 1979; McPherson *et al.* 2001, Hipp *et al.* 2011). Since proximity increases the probability of informal talks and social meetings (the 'office water cooler effect'), it is particularly valuable for the exchange of tacit and sensitive information. Face-to-face communication remains crucial, despite the rapid decrease of transport cost and the improvement of communication technologies that essentially favor the transmission of codified information (Liben-Nowell *et al.* 2005; Mok *et al.* 2010, Onnela *et al.* 2011). These general principles apply to VPOs as well, since they must often assess the advantages and disadvantages of conducting attacks in a distant location. Distance constrains attack locations in two ways. First, when attacks involve the same people or resources, these must be transported from one location to another, which takes time and costs money. Second, a distant location imposes transaction costs: unfamiliarity with the physical and social terrain, different languages, and so on.

Borders are one important aspect of the effect of distance. As Engel and Rogers (1996) and Borraz *et al.* (2016) showed, borders introduce price distortions that are equivalent to adding an extra distance between locations. Borders also limit social exchanges, even when people use social media (Takhteyev *et al.* 2011; Lee *et al.* 2011) and are a major impediment to labor market integration, despite formal agreements that promote the mobility of labor (Bartz and Fuchs-Schündeln 2012). In addition to hindering the mobility of goods and people, borders also have strong effects on political violence, which often occurs at the sub-national level (Dowd 2016). The effect of borders on the decision by a group to carry out attacks in more than one country can, therefore, be modelled as obstacles to be surmounted. The practical cause of the obstacle might be the overhead of the crossing, either overtly or covertly; differences in culture on the other side; or the increased risk associated with operating away from 'home turf', where, for example, it may be less obvious who can or should be bribed.

This paper presents a novel approach to border effects, which takes into consideration that, as the effect of borders increases, the resulting structure becomes less and less planar, and so a simple map representation becomes less and less accurate. The modification we suggest is instead to represent the locations where attacks have taken place as a network where the nodes are attack locations and the edges represent distances (increasingly modified) between them. In this context, the network representation has two advantages. First, the network need not be represented as a planar graph, so that we can model the effects of borders as distortions that stretch some edges, pushing the structure into the third (and possibly higher) dimensions. Second, the network can still be visualized by embedding it back into a geometric space of a small number of dimensions.

## 3. Geodesic and temporal distribution

The primary data source used in this article is the Armed Conflict Location and Event Data project (ACLED) dataset that records violence and protest in North and West Africa over the period 1997-2015 (acleddata.com). Rich data for each incident is available, including timing, groups participating as attackers and victims/targets, and location, both in terms of latitude and longitude, and by administrative district (Raleigh *et al.* 2010; Raleigh and Dowd, 2016). We restrict our attention to incidents that are clearly categorized as violent and fall into one of the following categories: Battle with and without change of territory; Riots and protests; Violence against civilians; and Remote violence. Attack locations for our purposes are at the granularity of local administrative districts.

## 3.1. The uneven geography of attacks in North and West Africa

As we would expect, statistics show that the distribution of the 29,272 attacks by 921 groups at 1831 locations is not random. The location where the most attacks took place is near Benghazi in Libya, where 1230 attacks are recorded. However, the mean number of attacks per location is 16, and the mean number of attacks for the least-attacked 1600 locations is only 4.7; so, the distribution is highly skewed. If we consider instead how many organizations have carried out attacks at each location, the highest score is a location near Tripoli where 60 different groups carried out an attack. However, the mean number of groups attacking at a given location is 3.7; so, again, the distribution is skewed.

These observations make it clear that this setting is far removed from conventional warfare where ground is taken and held; and that participants do not form large, stable blocs. Rather, interactions are fluid and consist of smaller, constantly shifting members and alliances. The decision about where to carry out an attack is constrained by two sets of factors: properties of the target, and properties of the attacking group. Clearly locations are attacked because of their properties: resources to be captured, propaganda value, and vulnerabilities. These properties can be analyzed by conventional risk analysis. The more difficult properties are those of the attacking groups, whose internal processes may be opaque, and who are fundamentally motivated to do the unexpected. However, such groups cannot attack at will – they are constrained by resources, broadly interpreted. Here we focus on the constraints imposed by distance (which matter in a non-urban environment) and the way borders distort distance.

Figure 1 shows the distribution of all attacks by latitude and longitude. Attack locations are not evenly distributed across the region. The main clusters of violence, by decreasing order of fatalities reported in the ACLED data, are principally located in Nigeria, Northern Algeria, Northern Libya, the Chad-Sudan border, and along the Gulf of Guinea. Nigeria is especially affected by violence in North and West Africa, with 50,144 fatalities, most of them resulting either from ethnic violence, fights to control oil production in the Niger Delta, or from attacks by Boko Haram. In West Africa, the border between Chad and Sudan remains a focus of conflict due to persistent fighting between the Sudanese government and rebels in neighboring Darfur. The portion of the Gulf of Guinea that extends from Abidjan to Banjul has suffered from a succession of civil wars in Ivory Coast, Liberia, Sierra Leone, and Guinea-Bissau.

In North Africa, Algeria has also been markedly affected by violence, principally due to activity by three organizations in conflict with the Algerian government: the Armed Islamic Group (GIA), the Salafist Group for Preaching and Combat (GSPC), and Al Qaeda in the Islamic Maghreb (AQIM). Violent Islamist groups were involved in 93% of the 12,050 fatalities in Algeria. With 12,610 fatalities reported, Libya is the third epicenter of violence, principally because of the overall political instability that followed the ousting of Col. Gaddafi in 2011 and the subsequent civil war. In comparison, the Sahel and Sahara regions are less immediately affected by violence, with the notable exception of Northern Mali where secessionist rebels and Islamist groups have opposed the government since 2012. More than 1,200 of the 2,761 victims of violent events reported in Mali from 1997 to 2015 died in an event involving one or several Islamist groups, including AQIM, Ansar Dine, the Movement for Oneness and Jihad in West Africa (MUJAO) and Al Mourabitoun. In Mauritania, where violent Islamist groups have also been active, the number of victims resulting from clashes with such groups is much lower, with 86 fatalities respectively, while in Niger, the number of victims (997) has experienced a rapid increase due to Boko Haram.

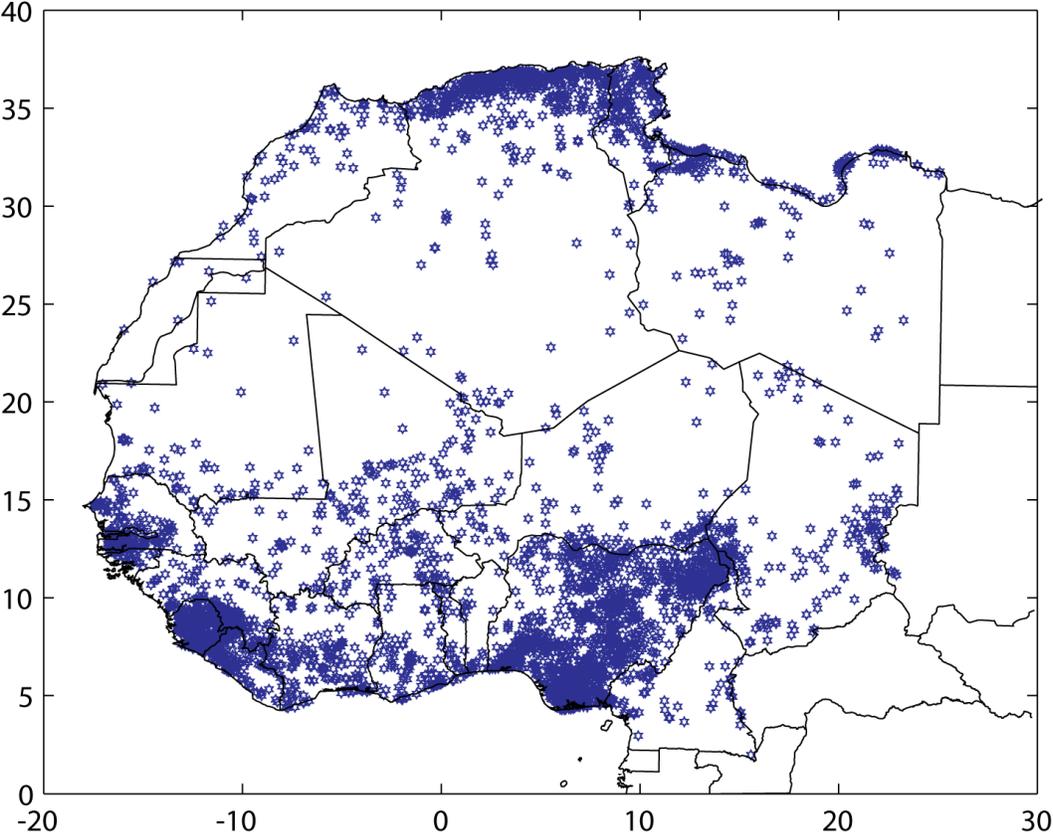

Fig. 1: The positions of all attacks by latitude and longitude. The Mediterranean coast can be seen at the top of the figure, the Atlantic coast on the left and the Gulf of Guinea on the lower side.

Some of the organizations whose attacks are captured in the dataset tend to focus on a single country. For example, several of the countries considered had civil wars over the period under consideration, and others had extended conflicts over presidential successions. However, it is still noteworthy how often organizations from one country carried out attacks in another. For example, the Islamist group AQIM, historically based in Algeria, has conducted numerous attacks in neighboring countries. Boko Haram, another VPO, is also responsible for attacking civilians and security forces in Niger, Chad and Cameroon. The transnational activity of VPOs has prompted many governments of the region to carry out attacks abroad: At the beginning of the 2010s, for example, the Mauritanian military carried out attacks in Mali to destroy military bases belonging to AQIM. More recently, Chad sent troop to both Nigeria and Cameroon to fight Boko Haram.

## 3.2. Temporal distribution

Over the last 20 years, North and West Africa have experienced episodic violence. As shown in Figure 2, the total number of fatalities was particularly high during the 1990s, the 'decade of despair' for Africa due to a rise in the number of conflicts on the continent that contrasted with the general decline observed elsewhere in the world at the end of the Cold War (Themnér and Wallensteen, 2014). Our data, which capture the last three years of the 1990s, highlight the high number of victims resulting from civil wars in Liberia (1989-1997 and 1999-2003), Sierra Leone (1991-2002), and Guinea-Bissau (1998-1999). The return of political stability to Sierra Leone and Liberia at the beginning of the 2000s coincides with the beginning of the first civil war in Ivory Coast (2002-2007). From the mid-2000s onwards, the rise of violent Islamist groups, shown in green, is evident. Apart from the peak in fatalities in 2011 due to the second civil war in Ivory Coast (2010-2012), the majority of the victims of recent conflicts are involved in clashes with violent Islamist groups, and their number is on the rise.

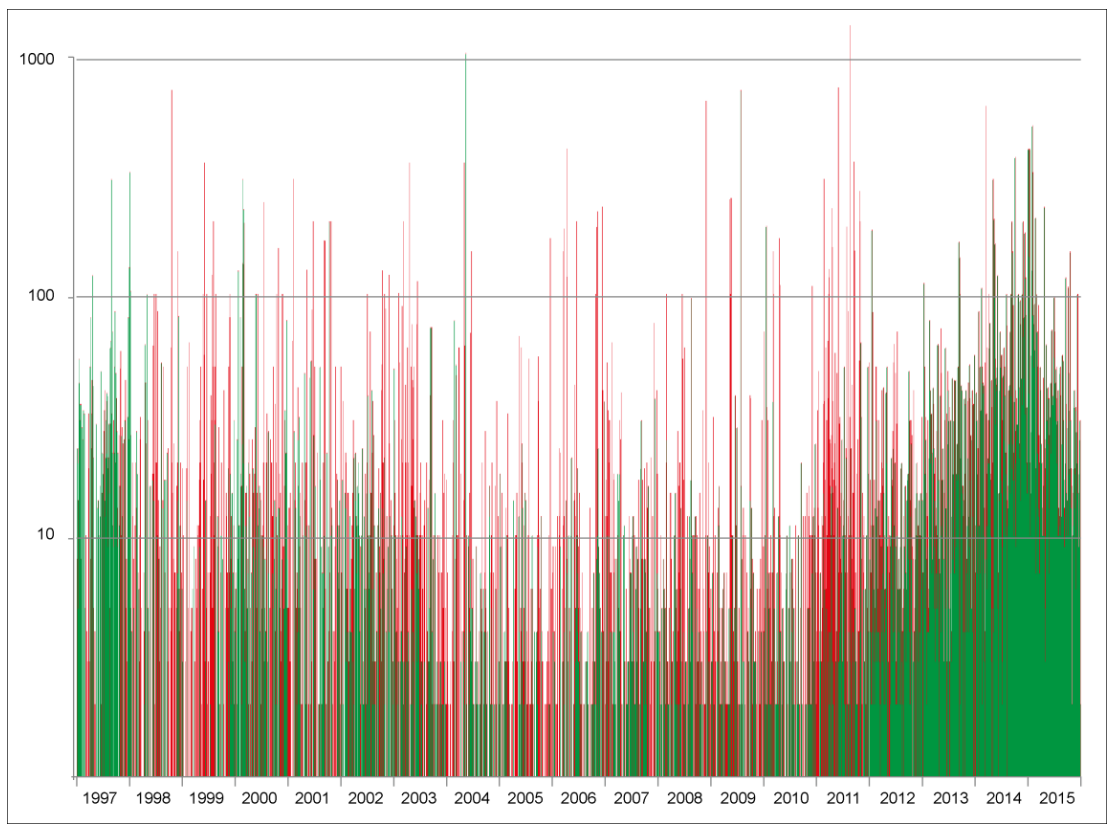

Fig. 2. Fatalities related to all groups (in red) and Islamist groups (in green), 1997-2015

## 4. Modelling with spectral embeddings

We begin with the graph derived from the location data and build an n × n adjacency matrix whose ijth entry is zero if the ith and jth nodes are not connected, or the weight of the edge connecting them if they are connected. If the edges are undirected, then the ijth and jith entries are the same, and the matrix is said to be symmetric. The matrix has 1831 nodes representing the attack locations, fully connected by edges whose edge weights are the geodesic (great circle) distances between them. These distances were calculated from the latitudes and longitudes using the haversine function.

## 4.1. Spectral embedding of graphs

To generate a networks whose nodes are administrative regions, we use spectral embedding (Spielman 2011; von Luxburg, 2007), a mathematical technique with two main steps.

First, the adjacency matrix is converted to one of a family of Laplacian matrices. We begin by using the combinatorial Laplacian, L, given by the matrix equation:

L = D – A

where D is the matrix whose ith diagonal entry is the total edge weight of the edges connected to node i, with all of its other entries zeros. Since A is symmetric, so is L.

Second, an eigendecomposition of L is computed such that:

L = Q Λ Q'

where Λ is a diagonal matrix of eigenvalues, in decreasing order, Q is the eigendecomposition of L, and the superscript dash indicated matrix transposition.

If the network is connected, then the final eigenvalue is zero, and we ignore the final column of Q. The k columns preceding it can be interpreted as the coordinates of each node of the graph in a k-dimensional space. In other words, if we take the n rows of Q, and the k = 2 columns n-1 and n-2 of Q we can use these as coordinates to place points corresponding to the n nodes in a 2-dimensional rendering of the network.

This spectral embedding comes with strong mathematical guarantees that it is the most faithful representation of the network structure in k dimensions (Spielman, 2011). From an intuitive viewpoint, the effect of spectral embedding is to begin with the cloud of points representing the nodes of the graph in a space of dimension n-1, in which the distances between each pair can be represented exactly. The eigenvectors are then oriented along orthogonal directions for which the cloud has the greatest variation: the n-1$^{st}$ eigenvector along the direction of greatest variation, the n-2$^{nd}$ along a direction orthogonal to this with the next greatest variation, and so on. The embedding is, therefore, a collection of directions from which the view of the graph shows the most possible differentiation.

Before we apply this spectral approach to the attack location network, we must alter the way in which edge weights are used. The edge weights in a network should be large for nodes that are strongly connected and small for nodes that are weakly connected. Distances, however, are the exact opposite – locations that are far apart (and so weakly associated) have large values for their mutual distance. The edge weights need to be inverted so that close locations have large weights and *vice versa*. There are several ways to do this, but we choose to subtract each distance from 1.1 × the longest distance between any pair of locations in the network.

In the figures that follow, the nodes of the network are color-coded by the countries in which they are located. Figure 3 shows an embedding of the attack locations based purely on geodesic distance between all pairs. The difference between this and Figure 1 (based on position) is that the network of locations draws them closer together when attacks happen close together in space. In other words, hotspots get hotter. National borders can be seen when the locations are color-coded by country, but would be much less obvious if they were not. This indicates that there are few differences in the locations of attacks along the Gulf of Guinea, or along the Mediterranean coast; but Burkina Faso and Mali show clear separations from their southern neighbors; and there is a strong separation between attacks in the Mediterranean countries and those farther south. Mauritania, and to some extent Algeria, are bridges between these two regions.

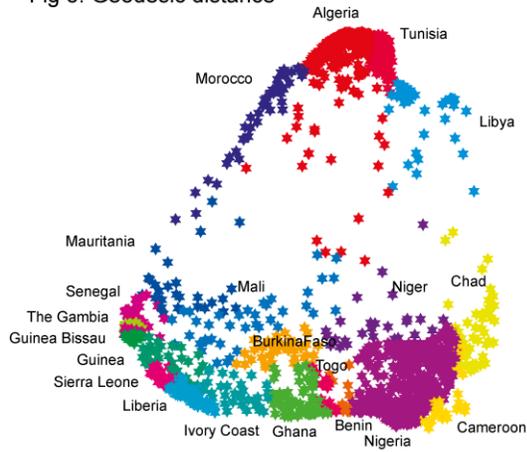
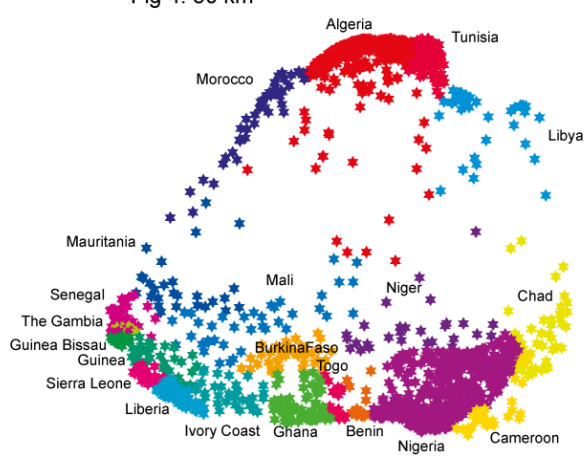
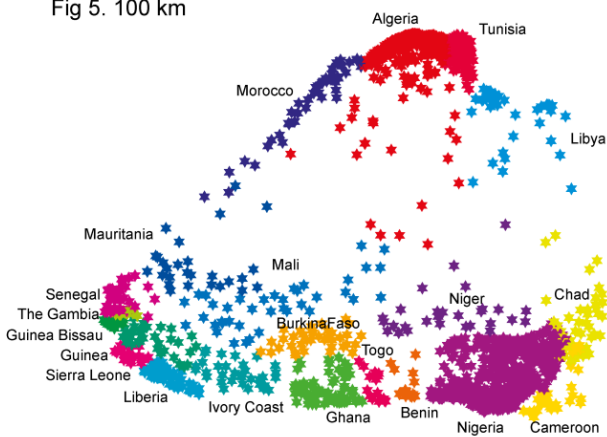
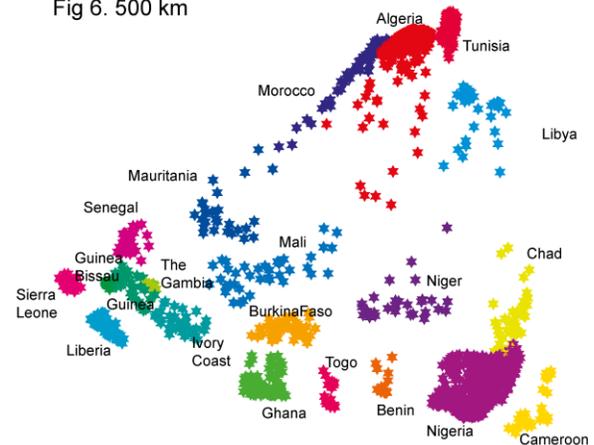

Fig. 3-6: Spectral embedding based on geodesic distance, with borders modelled as equivalent to distances of 50km, 100km, 500km

## 4.2. The effect of linear (additive) border costs

We now examine the effect of the presence of borders as it might enter into the calculus of a group planning its next attack. A simple way to model the cost of crossing a border is as an increased distance between origin and destination. For example, the addition of 100km to account for crossing a border captures a delay of, say, five hours (assuming typical speeds of 20km/h for travel) caused by the overheads of crossing.

We first compute the number of borders that must be crossed to pass between all pairs of the 21 countries we consider. This calculation was done based on the great circle distance between a median point in each country but when such a route would have required crossing many borders and a slightly longer route would have required crossing many fewer, the lower number of border crossings was used. For example, a direct path from Sierra Leone to Niger passes through Guinea, Ivory Coast, and Burkina Faso, but a path through Guinea and Mali crosses fewer borders without adding much distance.

In a model where border crossings are modelled as artificial added distances, the effect of multiple crossings is linear, since crossing two borders is twice as expensive as crossing one. Given a distance equivalent for each border crossed, we add this distance to the edge weight associated with each pair of nodes before inverting distances as described above.

Figure 4 shows the embedding when borders are modelled as equivalent to an increased distance of 50 km between countries. The maximum distance between attack locations in the dataset is almost 5000 km, so this is a small distortion, and indeed the differences between Figures 3 and 4 are too small to see at this resolution. However, when this distortion is increased to 100 km, the situation changes. Figure 5 shows the resulting embedding. On the one hand, attack locations in different countries now begin to separate in the visualization, indicating that they have become less similar, especially along the Gulf of Guinea. On the other hand, the border between Algeria and Tunisia shows little change, indicating how similar attack locations in these countries are. When the effect of a border is increased to be equivalent to 500 km, Figure 6 shows that locations clearly separate by country. When the cost of a border crossing is as great as this, cross-border locations seem less similar, and locations within the same country, by contrast, seem more similar to one another.

From the perspective of a group leader at a particular location and considering the location for a next attack, these results suggests that the presence of a border has little influence until the potential overhead of crossing that border is at least equivalent to the costs of 100 km of intra-country travel.

### 4.3. Non-linear border costs

While an argument could be made for linear border costs, it seems more plausible that the perceived cost of crossing borders is non-linear. For example, suppose that the probability of interdiction at any given border is 20%. Then the probability of interdiction when crossing two borders is 36%, since there is an 80% chance of successfully crossing the first border, and an 80% chance of success at the second border, so the probability of crossing both successfully in sequence is 0.8 × 0.8 = 0.64. Thus a group planning an attack two countries away should perceive it as substantially more costly than one in a neighboring country.

There are arguably two ways in which groups might frame non-linear border costs. On the one hand, a group with pan-national ambitions, such as AQIM, must exert its influence by carrying out attacks in countries that are far away from its center of influence. For such groups, a border-crossing cost might be appropriately framed in terms of success rate, and values as large as 95% would be necessary for them to succeed across states. On the other hand, a group whose interests are primarily domestic, such as Boko Haram, might regard borders as substantial impediments to their choices of attack locations, both because of the discomfort of operating in another country, and the reduced impact such an attack might have on their local agenda. For such groups, a much more substantial cost associated with crossing borders, perhaps 50%, might appropriately frame their calculus.

A border adjacency matrix for the 1831 locations was computed by setting the ijth entry to a given border success probability (between 0 and 1) raised to the power of the number of borders between the country of location i and the country of location j. If the probability of success is 0.95, then the result is $0.95^0 = 1$ for locations in the same country, 0.95 for locations in neighboring countries, $0.95^2 = 0.9$ for locations two countries apart and so on. If the probability of success is only slightly smaller, the effect becomes more pronounced. For a probability of success of 0.9 per border, the rate of success for crossing two borders sequentially is $0.9^2 = 0.81$ and for three, $0.9^3 = 0.73$.

## 4.4. Borders as a separate layer

Non-linear border costs cannot be represented as an addition to the representation of distances, because their impact depends on the particular pairs of locations being considered. Another way to incorporate their impact must be found. We do this by considering locations to be connected by two kinds of relationships: the obvious one based on how far apart they are, and the other by how many borders must be crossed on the path between them. Each of these two 'maps' looks roughly similar, but it is the differences between them that are most interesting. Our strategy is to align or reconcile them into a single representation that captures the effects of both kinds of distance simultaneously. We can then compute the closeness of any two locations taking full account of both relationships.

We have two adjacency matrices: one captures geodesic closeness and one captures border-based permeability. Both have the property that larger entries mean nodes are "closer". Our aim is to embed the network in such a way that the distance between points in the embedding reflects not only their physical distance apart, but also the effect of the number of borders that lie between them.

To do this we construct a new network and use a novel spectral embedding technique to embed it (Skillicorn and Zheng, 2012). The new network is built as follows: each node is replicated into two versions, a red version and a green version. The red versions of the nodes are connected using the adjacency matrix based on distances and the green versions are connected using the adjacency matrix based on border crossings. We can think of these two subgraphs as forming two layers. Now we connect each of the pairs of replicated nodes by new (say, blue) edges. Thus we have a single graph with red and green nodes, and red, green, and blue edges.

The question that now arises is how to assign weights to the new, blue edges. The larger these weights are, the more the two layers are forced to be "aligned". We can imagine that, in the larger graph, edges behave like springs that pull the nodes they connect with a force proportional to their edge weight. In each layer, nodes are pulled together based on their closeness (in the red layer) or country proximity (in the green layer), but they are also pulled together by how similar their "role" is in both layers at once. Note that the edge weights in the distance layer are much larger than those in the border layer.

To see how to assign weights to the blue edges, we convert the adjacency matrices to random walk matrices by computing the sum of edge weights in each row, and dividing the entries of each row by its sum. The matrix entries are now values between 0 and 1, which can now be interpreted as probabilities. The name "random walk" comes from imagining a walker who inhabits the network, constantly moving from node to node along the edges. At any given node, the walker chooses which node to visit next by choosing among the outgoing edges in proportion to their probabilities. The random-walker view of a network is quite elegant: for example, the proportion of time that a random walker spends at each node is a measure of its importance, since important nodes tend to be well-connected and so easy to visit regularly.

A variation of a random walk matrix, with better properties, is a *lazy* random walk matrix. Here the entries of each row are divided by twice the row sum, so that the entries sum to 0.5. The remaining 0.5 is placed in the diagonal position. The interpretation is now that the random walker makes decisions among the outgoing edges as before, but may choose, with probability 0.5, to remain at the current node for the next step.

We use this as a model to motivate the new $2n \times 2n$ random walk matrix. The entries corresponding to each subgraph are mapped to values between 0 and 0.5, the diagonals are left as zeros, and the remaining 0.5 probability is assigned to the blue edges between the layers. Thus, from a random walk perspective, a lazy random walker behaves in each layer as it would before, but can move from layer to layer with probability 0.5 on each step. Note that, because edge weights have been converted to transition probabilities, the differences in raw magnitudes between the layers have been normalized away.

The random walk matrix for the larger graph is bigger ($2n \times 2n$) but most of the extra entries in this matrix are zeros. The top left hand corner is the random walk matrix from the distances, the lower right hand corner is the random walk matrix of border crossing permeability, and the other two corners are diagonal matrices of the edge weights of the blue edges, and so mostly zeros. The cost of computing an eigendecomposition depends on the number of non-zero values in the matrix, so computing an eigendecomposition for the bigger graph is not much more expensive than computing one for each of the existing networks separately.

Now the standard spectral embedding algorithm can be used to convert this larger adjacency matrix to a Laplacian, compute its eigendecomposition, and embed the resulting graph in a 2- or 3-dimensional space. In this random-walk embedding, each location is represented by embedded red *and* green points. The distance between the two points corresponding to the same location (the length of the embedded blue edge) reflects how different their roles are from the perspective of distance and the perspective of borders. Locations for which these points are far apart are of particular interest.

Figure 7 shows the full embedding of the two-layer graph with border-crossing success probability set to 95%. The red versions of the nodes lie around the outside of the embedding, being "pushed" apart by the effect of borders; the green versions of the nodes lie further towards the center, pulled inwards by a relatively smaller effect of border crossing costs; and the blue lines indicate the magnitude and direction of the difference for each location.

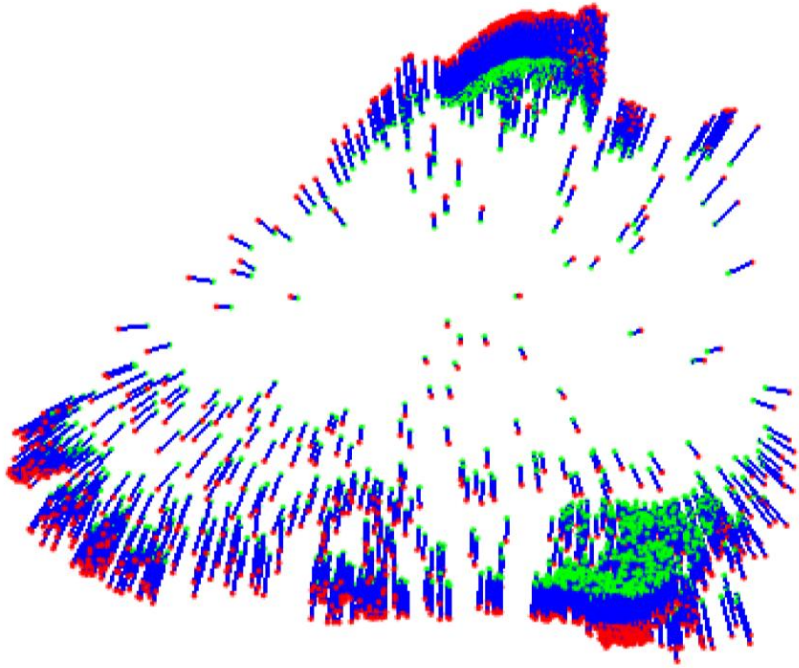

Fig. 7: Embedding of two-layer graph (red – locations based on distance, green – locations based on border crossings, blue lines – difference between the two). Borders push locations apart by amounts that depend on the global similarities.

Figure 8 shows the embedding of the locations based on distance, which represents our "red" nodes color-coded by the countries in which they are located as before. Comparing this figure to Figure 5, where borders were represented as equivalent to distances of 100 km, shows that the spread of locations is not dissimilar – but there is a greater spread from east to west, as expected given the number of borders along the Gulf of Guinea.

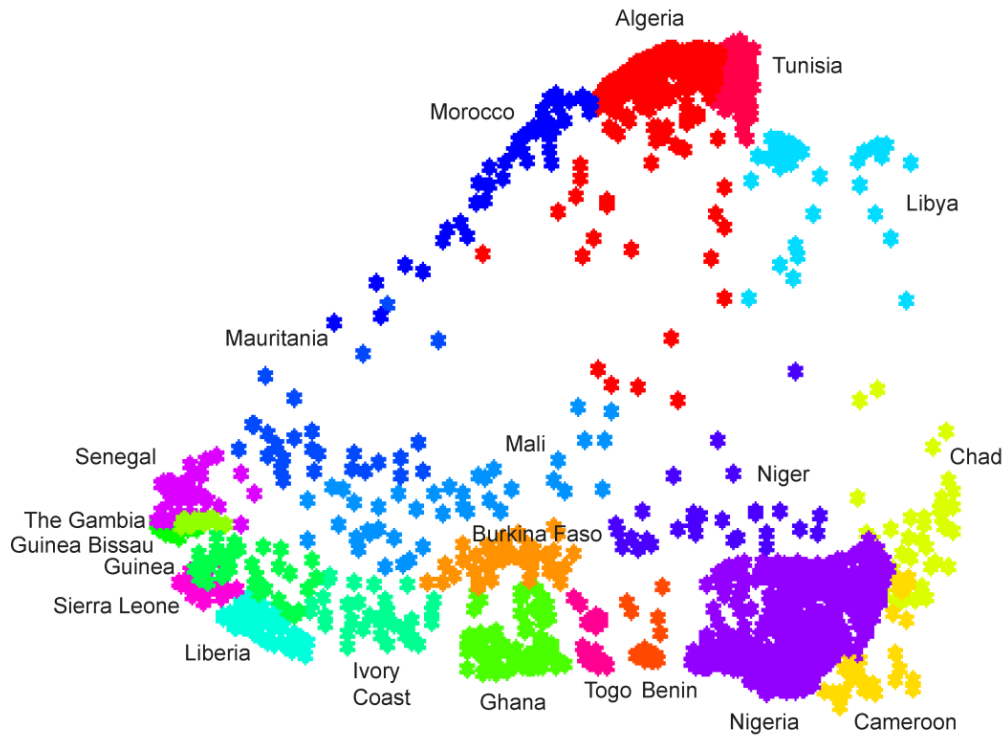

Fig. 8: Spectral embedding with non-linear border costs, crossing probability 0.95.

Figure 9 shows the difference between locations based on distance and based on border crossings. The "blue" lines, also color-coded by country, make it clear that the effect of borders is effectively to spread locations further apart from a virtual center in Southern Algeria, a fixed point where the distance to all other locations in North and West Africa is proportional to the number of borders that have to be crossed to reach them. This point corresponds to the commune of Bordj Badji Mokhtar, in Adrar Province, Algeria. Bordj Badji Mokhtar and the adjacent trading town of Al-Khalil in Mali have long been known for being a haven for arm and drug traffickers and a central node in the transnational network that connects local Tuareg tribes with the Algerian military and secret police (Scheele 2012). Between 2008 and 2014, Bordj Badji Mokhtar has been the theatre of repeated clashes between Algerian governmental forces and Islamist groups such as AQIM and MUJAO. Other examples of particular remote but 'rational' locations for crossing numerous borders include the small Algerian town of Tin Zaouten, which has experienced several major conflicts between Algeria, AQIM and the MUJAO since the beginning of the 2010s, and the Salvador Pass located at the crossroads between Niger, Libya and Chad, more than 1200 km from the nearest capital city, where traffickers and violent Islamist groups have clashed with French and Nigerien military forces stationed in the area.

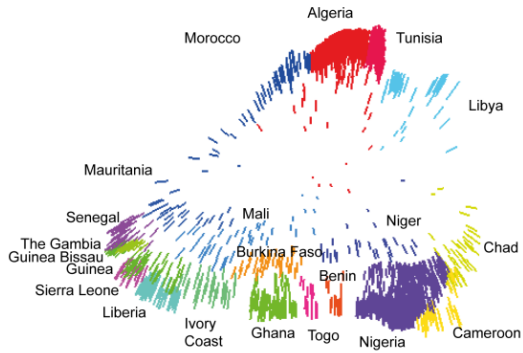

Fig 9. Crossing probability: 0.95

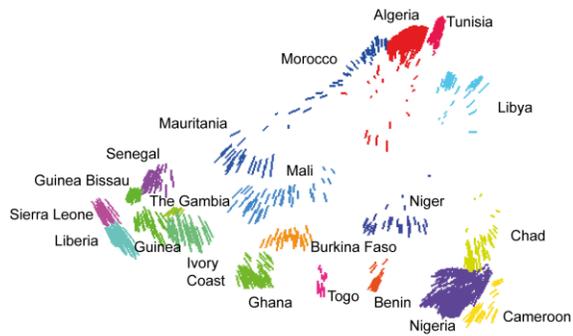

Fig 10. Crossing probability: 0.80

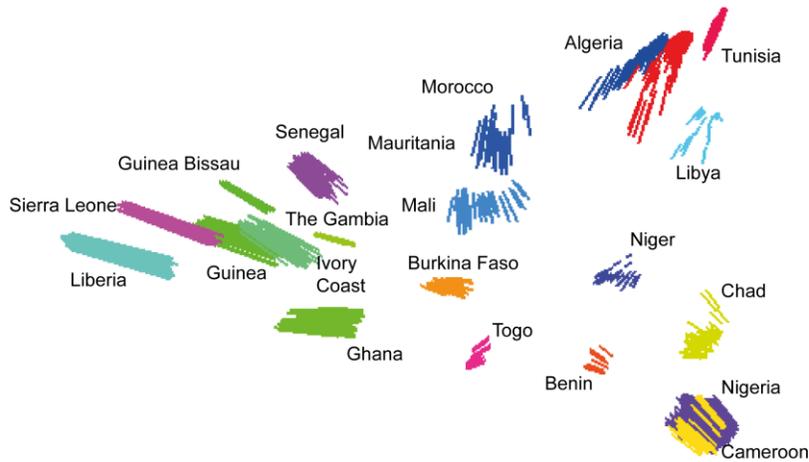

Fig 11. Crossing probability: 0.50

Fig. 9-11: Difference between embedded positions based on distance and border crossings: crossing probability 0.95, 0.80, 0.50

Figure 10 shows a similar figure but with the border-crossing success rate reduced to 80%. With this assumption, the distortion introduced by borders is quite different. Locations in the north and center of the region show the same radial distortion, in which locations appear further apart than they are because of the presence of borders. But, for the countries in the southwest and southeast, the distortion introduced by borders is oriented orthogonally to the distortion previously seen. For example, Sierra Leone and Liberia "push" one another apart rather than being influenced by distant Algeria and Tunisia; and Nigeria and Cameroon show a similar pattern.

Figure 11 shows what happens when the probability of successfully crossing borders is reduced to 50%, reflecting the mindset of groups with primarily local agendas. Distortions caused by borders are almost completely local, depending primarily on a few near neighbors. Because of the roughly triangular shape of North and West Africa, the net effect is that most of the distortions align toward the center but, since the probability of crossing a substantial number of borders drops quickly to a small value, the countries are only weakly connected. Note that Cameroon sees the rest of the region through the lens of Nigeria, from which it is now indistinguishable. This situation reflects well the increasing interdependence between the two countries since the jihadist group Boko Haram, historically active in Nigeria, spread to adjacent countries in 2014.

## 4.5. Time sequence as a separate layer

So far, we have ignored the dimension of time. The attacks carried out in the name of a single group might have involved separate subgroups, and so be more or less independent. However, it is natural to consider the sequence of attacks by each separate group. Unless subgroups of the group act completely independently, there are some constraints on the sequences of attacks, and these may offer insights into its constraints or strategy. For example, if the same subgroup carries out successive attacks, they must travel from one location to another, perhaps taking or gathering materiel; and perhaps crossing borders as well. Even if the attacks are not carried out by the same individuals, the time sequence reflects, at some level, strategic thinking by the group's leadership.

From the ACLED dataset we extract a third adjacency matrix connecting successive attacks by the same group by a directed edge of weight 1. If the ijth entry is 1, then the jith entry is unlikely to be, so this matrix is typically asymmetric.

Our analysis include 10 groups: Boko Haram and 9 other Islamist groups affiliated with Al Qaeda, who share a common historical and ideological background and form several components of a single, flexible network: Al Qaeda, Ansare Dine, AQIM, GIA, Al Mourabitoune, GSL, GSPC, MUJAO, and Those Who Signed in Blood (Walther and Leuprecht, 2015). One should however note that AQIM has historically occupied two very distant regions: the Kabylie region in northern Algeria, where its 'national' emir Abdelmalek Droukdel is still supposed to reside, and the Sahel-Sahara region where several sections (*katibas*) have developed since the mid-2000s. Over the years, major divergences of opinion have developed between the 'national' leadership in Kabylie and its Saharan section leaders (Walther and Christopoulos, 2015), who have repeatedly been criticized for their autonomous and often criminal operations (Associated Press, 2013; Wojtanik, 2015). Furthermore, the two 'groups' are often unable to communicate freely due to the distance and counter-terrorism operations of the Algerian military. Mapping the chronological activity of both AQIM's Kabylian leadership and Saharan sections would give the impression that much of the attacks span the Saharan desert, which is probably not realistic. To solve this issue, we have considered separately all violent events located south of In Amenas in Algeria.

Figure 12 shows the locations' embedded distance as in Figure 3, but overlaid by black edges connecting sequential attacks by the same organization (that is, whenever there is a 1 in the newly constructed directed adjacency matrix). The map confirms the pan-regional ambition of the 9 Islamist groups affiliated with Al Qaeda, who conducted attacks from Mauritania to Chad, often across borders. Their attack patterns diverge greatly from those of Boko Haram, the majority of whose violent attacks took place within Nigeria itself (Dowd, 2015; 2016).

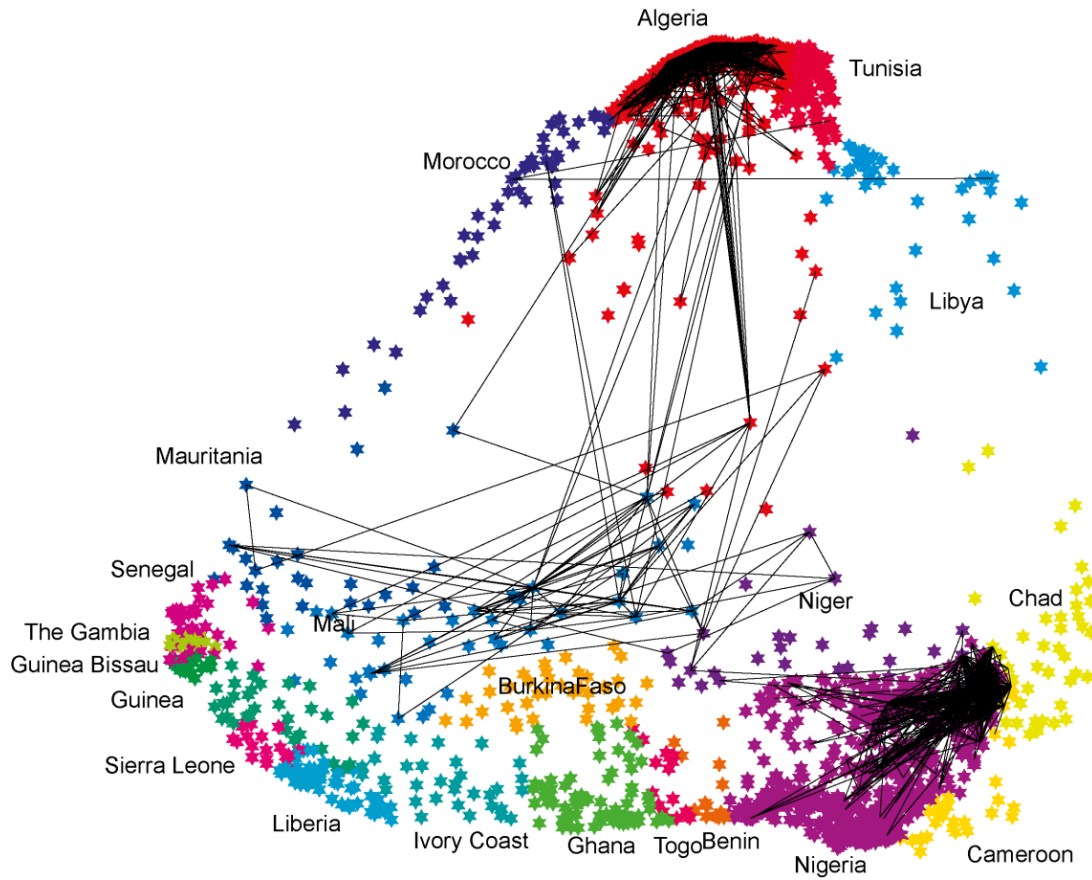

Fig. 12: Spectral embedding as in Figure 8, overlaid by lines connecting sequential attacks by the same group (for 10 groups with pan-national intentions)

We now want to extend the embedding to include the time-sequence structure. We do this by extending the layered model to three layers: one representing geodesic closeness, one representing border permeability, and one representing sequence in time.

The fact that the edges are directed in the new third layer introduces three complications to the embedding process. First, in an undirected network, the importance of a node is proportional to the total edge weight of the edges connected to it. This is no longer true for a directed network – a node may have many heavily weighted incoming edges, but it may not be important if these upstream nodes themselves are hard to reach. Calculating importance requires analyzing the entire graph, rather than being a local property. Second, a random walker can become trapped at a single node that only has incoming edges (but this is easy to detect), or in a region that collectively has no outgoing edges (and this is expensive to detect). Third, a random walk matrix is not necessarily symmetric, which the standard embedding algorithm requires.

A solution to these problems was developed by Chung (2008), but it has a number of drawbacks. Instead, we use a newly developed approach that models edge direction by replicating each node into an outgoing version and an incoming version, and connecting these in the obvious way by *undirected* edges. This reduces the problem to embedding an undirected graph at the expense of adding more layers. As before, new edges have to be added between the versions of the same nodes (Skillicorn and Zheng, 2014). This approach has been validated against the known social structures of the Florentine families in the time of the Medicis (Skillicorn and Zheng, 2014), and has been applied to understand the social networks of criminal networks (Skillicorn *et al*., 2015).

The construction is as follows: we combine three layers, the border accessibility layer (with success probability 0.95), the geodesic distance layer, and the time sequence layer. In the previous construction we converted each layer into a random walk matrix, divided the total edge weights incident at each node in half, and allocated half to edges that remain in the layer (proportional to their original values) and half to the edge to the other layer. We cannot follow this strategy for the three-layer graph because the directed adjacency matrix cannot be converted to a random walk matrix. However, we do use the same intuition: the total outgoing edge weight incident at each node should be divided in half, with half remaining within the layer, and half allocated to the edges to other layers. There are now two other layers, so the amount allocated to other layers is split equally between them.

The edge weights in the border accessibility layer are between 0 and 1; the edge weights in the geodesic layer are between 0 and ~4500, and the edge weights in the time-sequence layer are between 0 and 16. If we use this approach directly, edges from the geodesic distance layer to other layers will dominate, since their weights are so much larger than those of the other two layers. To compensate we must apply a normalization to each of the subgraph adjacency matrices to make the magnitudes of the edge weights comparable between the layers. We do this by dividing the entries for each layer by the mean of the non-zero entries in that layer. Almost all non-zero values are close to 1, with the exception of larger entries in the sequence layer.

A further complication arises because of the extreme sparsity of the sequence layer; in comparison, the other two layers are fully connected. If the sequence layer were to be embedded by itself, there would be two strong clusters with a weak connection between them, and many isolated nodes. The isolated nodes would be embedded at the origin, and the two clusters in a dumbbell shape. What actually happens in this layer can be regarded as altering this layer by making it fully connected by small constant weights (which is very roughly what the other two layers can be considered to do). The net effect is that nodes are placed more centrally (from the point of view of the sequence layer) depending on how weakly they are connected to other nodes in the layer. Nodes with no connection in the layer are embedded close to the center, nodes with weak connections in the layer are embedded further out, and nodes that are strongly connected in the layer are embedded furthest out. This is, of course, the inverse of what we would expect – important nodes being embedded centrally. The solution is to further normalize the edge weights in this layer by making the sum of the edge weights constant – so that nodes without any connections in the layer are given a heavily weighted self-loop, nodes with weaker connections in the layer a less-weighted self-loop, and so on. The result is a three-layer structure, with added *directed* edges between copies of the same node in the different layers.

We now replicate each layer into two (further) sublayers, one to which the outgoing edges are attached, and one to which the incoming edges are attached. All edges are now undirected; so, we can use the standard embedding (on a 6n × 6n matrix whose entries are almost all zeros). The newly replicated copies must also be connected to one another by edges whose weights depend on the incident edge weights of the pair. This construction is intricate but essentially straightforward (Zheng and Skillicorn, 2015).

The resulting embedding is shown in Figure 13. It is clear that Boko Haram attacks are local to Nigeria, with occasional incursions into Chad and Cameroon. There are several locations that are the site of repeated attacks. The attacks by Islamist insurgents affiliated with Al Qaeda are concentrated in Northern Algeria, but another nexus of attacks can be observed in Southern Algeria, Mali, Mauritania, and Niger. These second group of attacks are much more widespread physically. Note that, as locations attacked sequentially appear more similar, locations where this does not happen spread further apart in the embedding.

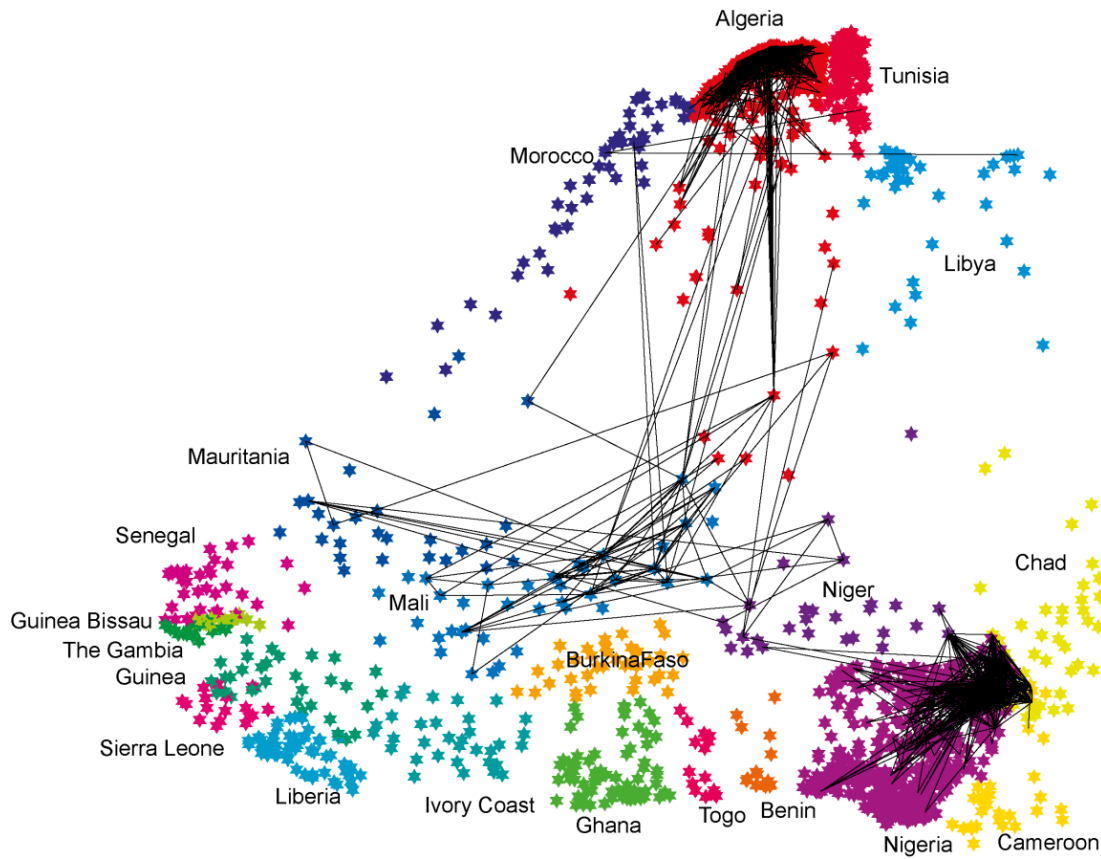

Fig. 13: Spectral embedding of the 3-layer network, with locations attacked sequentially appearing more similar. Note the difference from Figure 12, where sequence plays no role in embedded position.

## 5. Conclusion

The aim of this article was to provide a dynamic description of the structure of conflict in North and West Africa across time. While necessarily limited in the causal inferences that can be drawn, our findings significantly advance the state of knowledge in network science and conflict studies.

First, the paper expands our understanding of the way structural data can be extended into the analysis of conflict through the novel application of spectral embedding techniques to network science. We have shown how newly developed extensions to spectral embedding techniques (with typed edges, and directed edges) that have been previously applied to conventional social networks, with humans as nodes and relationships as edges, can be extended to novel social networks in which the nodes are locations and edges represent perceived distances.

Second, the paper prods us to think differently about the nexus of space and conflict: commonly we think of networks as a function of place; instead, this paper inverts this convention to think of place as a function of networks. Place is no longer simply a physical location – its positional and strategic importance define it, and these change as the movements that give rise to networks change. Our study of political violence in Africa is an application of ideas from social networks to networks of a different kind, reflecting the fact that place is a human construct as well as a physical one. An interesting finding of our work is to show that some of the most violent places in the region are situated virtually in the middle of nowhere. In the extreme north of Mali or in the northeastern reaches of Niger, violent events create ephemeral places of action hundreds of miles from any inhabited areas. Instead of thinking of conflicts as a function of place *per se*, we can now think of conflicts as a function of movements. Since movements, unlike place, are not fixed, strategic consideration can now be given to ways to influence, alter, or disperse some movements while generating and encouraging others (Retaillé and Walther, 2013). We have long known that conflict is dynamic; this paper posits a method to model that dynamism, one that makes it possible to respond to conflict and violence in terms of strategic consideration of movement rather than simple spatial coordinates.

Finally, spectral embedding of these networks of places provides an insight into the potential mindset of group commanders as they consider their next actions. Geodesic distance obviously plays a role, but other constraints are also significant. Borders are one such constraint. We have modelled the effect of borders from two perspectives: that of a group with pan-national aspirations, and that of a group with more local aspirations, and shown that the mental landscapes produced are very different. In other words, we can show that strategic intention affects the framing of the transaction costs imposed by crossing borders and resource constraints of operating across vaster distances, and so affect the calculus of a group's leadership. Constraints as simple as habits also play a role: there are some pairs of locations where attacks took place at the same pair of locations in sequence 16 times.

Our findings have implications for conflict prevention and early intervention in the region. For example, consider the practical problem of anticipating, after an attack by a particular group, the probable timing and location of the next attack. Without extrinsic information, the conventional approach is simply to draw concentric circles around the location of the current attack, and assign a reduced probability the more distant each location. However, if we know that borders represent an impediment to movement (whose cost we can estimate), or that habits play a role, or that target groups are themselves mobile, then these contours of probability should be warped, concave in shape where movement in that direction is hard, and convex when movement is easy (even mentally easy, as in a habit). Our technique forgoes this complex, *ad hoc* manipulation, and warps the space instead, so that locations that are easier to reach (physically, border-crossing wise, or strategically) are embedded closer together, and locations that are hard to reach are embedded further apart. As a result, concentric circles again demarcate regions that are equally likely to be the site of the next attack. In other words, a concentric circle drawn on any of the embeddings in this paper represent locations at similar strategic distance from its center, given the cost assumptions associated with that figure.

Myriad other measures potentially modulate the role of distance in projecting the "next" place. The varying difficulty and desirability of border crossings for migrants into and within Europe is one example: some countries are attractors and so the distance to them may seem shorter than they are in the mental calculus of migrants, but some borders are harder to cross, making distances that involve them seem longer than they are. The ability to incorporate these multiple modulating factors with distance, to create a mental map that combines all of these varying criteria into a framework that migrant populations might hold and informs their behavioral logic enables both insight and strategic reaction in what might otherwise appear as wholly unpredictable settings.

Our approach to political violence gets us closer to a prototype of predictive modeling of conflict of the sort that is widely used by urban police to anticipate crime. We can now model some of the decision-making calculus that informs leaders' decisions with respect to potential target sets and resource allocation by weighing the costs and benefits of such decisions in the light of transaction costs, such as those imposed by borders. Here we have treated all borders as equal barriers, but the methodology would also allow more detailed modelling, for example the efforts by the Mauritanian government to enhance their border security since 2011 by increasing patrols and working with local tribes.